\documentclass{ws-ijmpd}
\usepackage{epsfig}
\usepackage{epstopdf}
\usepackage{color}

\begin{document}

\title{Cosmological models with a Hybrid Scale Factor}

\author{S.K. Tripathy}
\address{Department of Physics, Indira Gandhi Institute of Technology, Sarang, Dhenkanal, Odisha-759146, India\\ tripathy\_sunil@rediffmail.com}

\author{B. Mishra}
\address{Department of Mathematics, Birla Institute of Technology and Science-Pilani, Hyderabad Campus, Hyderabad-500078, India\\ bivu@hyderabad.bits-pilani.ac.in}

\author{Maxim Khlopov}
\address{National Research Nuclear University, MEPHI (Moscow Engineering Physics Institute), \mbox{Moscow 115409, Russia}; CNRS, Astroparticule et Cosmologie, Universit\'e de Paris, F-75013 Paris, France \& Institute of Physics, Southern Federal University, 344090 Rostov on Don, Russia\\ khlopov@apc.in2p3.fr} 

\author{Saibal Ray}
\address{Department of Physics, Government College of Engineering and Ceramic Technology, Kolkata 700 010, West Bengal, India\\ saibal@associates.iucaa.in}
	
\maketitle	

\abstract{In this brief review, we present some cosmological models with a Hybrid Scale Factor (HSF) in the framework of general relativity (GR). The hybrid scale factor fosters an early deceleration as well as a late time acceleration and mimics the present Universe. The dynamical aspects of different cosmological models with HSF in the presence of different matter fields have been discussed.}


\section{Introduction}
The relative expansion of the Universe is parametrized by a dimensionless scale factor $\mathcal{R}$. This is a key parameter in Friedman equations and also known as the cosmic scale factor or Robertson Walker scale factor. In the early stages of the Big Bang, most of the energy was in the form of radiation and that radiation has a dominant influence on the expansion of the Universe. Later, with cooling from the expansion, the roles of matter and radiation changed and the Universe entered into a matter dominated era. Recent observational results suggest that we have already entered an era dominated by dark energy (DE). But investigation of the roles of matter and radiation is most important for a good understanding  of the early Universe. One should note that, the effective energy density of the Universe is usually expressed in terms of the scale factor. Also, the dynamics of the Universe is assessed through an equation of state parameter $\omega$ usually defined as the ratio of the pressure $(p)$ to the energy density $\rho$. Within the purview of GR using a flat Friedman model and assuming a constant equation of state parameter $\omega$, the energy density relates to the scale factor as $\rho \sim \mathcal{R}^{3(1+\omega)}$. The evolution of the scale factor is a dynamical question, determined by the equations of general relativity, which are presented in the case of a locally isotropic, locally homogeneous universe by the Friedman equations. In a flat Universe model comprising only a perfect cosmic fluid, we may have the scale factor as $\mathcal{R}(t)\sim t^{\frac{2}{3(1+\omega)}}$. For a radiation dominated Universe, the scale factor behaves like $\sim t^{1/2}$, whereas for a matter dominated era, the scale factor behaves like $\sim t^{2/3}$. In the dark energy dominated era, we may have a De Sitter type expansion with the scale factor behaving like an exponential function such as $\mathcal{R}(t)\sim e^{H_0t}$. 

From the above simple discussion, we may infer that the equation of state parameter becomes $\omega=\frac{1}{3}$ for radiation era, $\omega=0$ for matter dominated era and for a DE dominated era we may have $-\frac{2}{3}\leq \omega \leq -\frac{1}{3}$. On the other hand, different values of the equation of state parameter have been extracted from observational data in recent years. DE models with a cosmological constant ($\Lambda$CDM model) predicts the equation of state parameter as $\omega=-1$ and for quintessence models, we have $\omega > -1$. However, as inferred from recent observations, there may be possibility of a phantom field dominated phase with $\omega <-1$ \cite{Tripathi2017}. The 9 year WMAP survey suggests that $\omega=-1.073^{+0.090}_{-0.089}$ from CMB measurements and $\omega=-1.084\pm 0.063$ in combinations with Supernova data\cite{Hinshaw13}. The Supernova cosmology project group have found that $\omega=-1.035^{+0.055}_{-0.059}$ \cite{Amanullah2010}. From a combined  analysis of the data sets of SNLS3, BAO, Planck, WMAP9 and WiggleZ, Kumar and Xu  constrained the equation of state parameter as $\omega=-1.06^{+0.11}_{-0.13}$ \cite{Kumar2014}. Moreover the recent Planck 2018 results constrained $\omega=-1.03\pm 0.03$ \cite{Planck2018}.  

It has been confirmed from a lot of observations that, the Universe is  in a state of accelerated expansion at least at its late phase of evolution \cite{S.per 1998,A.G 1998,R.Knop 2003,A.G 2004,A.G. 2007,D.N 2007,D.J 2005,M.Sulli 2011,N.Suzuki 2012,C.R 2004,S.W 2004,S.P 2004,Cole2005}. Also, it is believed that, the Universe was decelerating in the past prior to entering into an accelerated phase. The deceleration is mostly due to a matter dominated phase of the Universe. After the Universe enters into a DE dominated phase, the scale factor increases exponentially. At the very beginning after the phenomenal Big Bang, the Universe is also believed to have undergone an exponential expansion. This phase is popularly known as the inflationary phase. After this inflationary phase, the Universe evolves to decelerate and undergoes a transition from a decelerated phase to an accelerated phase at its late phase. The transit epoch, when the Universe flips from a decelerated phase to an accelerated one corresponds to a  transit redshift $z_{da}$ which is believed to be an order of 1 i.e $z_{da} \sim 1$. The transition redshift has been constrained to be $z_{da}= 0.82\pm 0.08$ (Busca \cite{Busca13}), $z_{da}=0.74\pm 0.05$  (Farooq and Ratra  \cite{Farooq13}), $z_{da}= 0.7679^{+0.1831}_{-0.1829}$ (Capozziello et al. \cite{Capo14}), $z_{da}=0.69^{+0.23}_{-0.12}$ (Lu et al. \cite{Lu11}), $z_{da}=0.4\pm 0.1$ ( Moresco et al. \cite{Moresco16}). While Reiss et al. derived  kinematic limits on the transition redshift as $z_{da}= 0.426^{+0.27}_{-0.089}$ \cite{Reiss07}, in a recent work, Goswami et al. obtained a constraint as $z_{da}=0.73$ \cite{Goswami2021}. If at all, the present Universe follows the DE models, then, this transition redshift $z_{da}$ may be considered as an important cosmological parameter. 

The dynamical aspects of Universe may be modelled in many ways. Usually, for a given gravitational theory such as GR, the field equations are set up assuming a cosmic fluid comprising some matter fields and then for an assumed equation of state (may be dynamically varying), the field equations are solved to obtain the cosmic dynamics. Other ways to investigate the cosmological issues may include, the assumption of a cosmic dynamics through an adhoc scale factor and then the relationship between the energy density and pressure is obtained. The later modelling method provides an opportunity to obtain the equation of state parameter for a dynamically evolving DE phase. Here we require a suitably chosen scale factor that mimicks the present Universe. 

In fact, there are several scale factors available in the literature. A power law form $\mathcal{R}(t)\sim t^{n}$, $n$ being a constant  and an exponential expansion form $\mathcal{R}(t)\sim e^{\alpha t}$, $\alpha$ being a positive constant are well known. Bouncing scale factors providing a bouncing scenario to avoid the singularity problem occurring at the initial epoch have different structures than these two forms. However, these scale factors provide a constant deceleration parameter $q=-\frac{\mathcal{R}\ddot{\mathcal{R}}}{(\dot{\mathcal{R}})^2}$. But, consequent upon the recent finding concerning the late time cosmic speed up, we require a signature flipping deceleration parameter having positive values at early times and negative values at late time. Such a deceleration parameter may be obtained from a hybrid scale factor \cite{Mishra15} having two factors: one behaving as a power law and the other behaving as an exponential law of expansion. While the power law factor of the HSF dominates at an initial epoch, the exponential factor dominates at late time to provide a suitable explanation for the transitioning Universe.

In the present review, we wish to present the role played by the hybrid scale factor in obtaining the dynamical behaviour of the Universe. We have considered diversified space times and different matter fields in the framework of general relativity to justify that, HSF can be a good alternative as an adhoc scale factor for investigating background cosmologies. The article is organized as follows: In Sec. 2, we present a brief idea about the hybrid scale factor. In Sec. 3, some viable cosmological models have been discussed where we have employed HSF. At the end in Sec. 4, we summarize our results.

\section{Hybrid Scale Factor}
The hybrid scale factor may be expressed as \cite{Mishra15,Mishra18,Mishra2018,Mishra21,Tripathy2020}
\begin{equation}
\mathcal{R} =e^{\alpha t}t^{\beta},\label{eq:1}
\end{equation}
where $\alpha$ and $\beta$ are positive constants. As it will be described, a cosmic transit from early deceleration to late time acceleration can be obtained using the hybrid scale factor. It is certain from \eqref{eq:1} that, the power law behaviour dominate the cosmic dynamics in early phase of cosmic evolution and the exponential factor dominates at late phase. When $\beta=0$, the exponential law is recovered and for $\alpha=0$, the scale factor reduces to the power law. The Hubble parameter and the deceleration parameter for the hybrid scale factor can be obtained as
\begin{eqnarray}
H &=&\alpha+\frac{\beta}{t},\\
q &=& -1+\frac{\beta}{(\alpha t+\beta)^2}.
\end{eqnarray}

Similar expansion law has already been conceived earlier. In a recent work, Tripathy \cite{SKT2014} has used a more general form of such hybrid Hubble parameter has been considered with the form $H=\alpha+\frac{\beta}{t^n}$. The present HSF model is a special case of the scale factor considered in Ref. \cite{SKT2014}. One may note from the expression of the deceleration parameter that, at an early phase of cosmic evolution, the deceleration parameter becomes $q=-1+\frac{1}{\beta}$ and at late phase of cosmic evolution it approaches to $-1$. In order to obtain a signature flipping behaviour of the deceleration parameter having positive values at an early time and negative values at late times, it is required to adjust the value of the parameter $\beta$ in the range $0<\beta<1$. Since, at the cosmic transit epoch, the deceleration parameter vanishes, we can infer that, the cosmic transit occurs at a time $t_{da}=-\frac{\beta}{\alpha}\pm \frac{\sqrt{\beta}}{\alpha}$. The negativity of the second term leads to a concept of negative time which appears to unphysical in the context of Big Bang cosmology and therefore, we should have $t_{da}=\frac{\sqrt{\beta}-\beta}{\alpha}$. It is possible to study different issues in cosmology concerning a transitioning Universe with a behaviour of early phase deceleration and late phase acceleration by employing an assumed dynamics through the hybrid scale factor. The parameter $\beta$ has a defined range of $0<\beta<1$. From an analysis of the behaviour of an anisotropic model, Mishra and Tripathy have constrained the parameter $\beta$ in a further narrow range $0<\beta<\frac{1}{3}$ \cite{Mishra15}. This range is required to get a positive time frame to have a decelerated Universe. The other parameter $\alpha$ may be constrained from a detail analysis of $H(z)$ data \cite{Tripathy2020}. Mishra et al. have investigated some anisotropic dark energy models in the framework of GR and have employed a hybrid scale factor to study the cosmic dynamics \cite{Mishra18}. In that work, they have used the range $0<\beta<1$ and constrained the value of of $\alpha$ in the range  $0.075<\alpha <0.1$ basing on recent observational constraints on transition redshift $0.4<z_{da}<0.8$. Particularly they have used two specific values of $\alpha$ namely $0.1$ and $0.075$. Through the behaviour of the deceleration parameter as deciphered from the hybrid scale factor, Mishra et al. have shown that the power law factor dominate at the early time and the exponential factor dominates at late time. The rate of transition of the deceleration parameter is obtained in that work to be faster for a higher value of $\alpha$. In a recent work of Tripathy et al. \cite{Tripathy2020}, the hybrid scale factor has been constrained from an analysis of the $H(z)$ data and four different HSF models have been proposed. However, accurate determination of the HSF parameters is essential to get a model more closer to the present Universe.

Usually, the cosmological models with dark energy components are distinguished through the use of two important diagnostic approaches: the determination of the state finder pair $\{j,s\}$ in the $j-s$ plane and the $Om(z)$ diagnostics. While the state finder pair involve third derivatives of the scale factor, the $Om(z)$ parameter involve only the first derivative of the scale factor appearing through the Hubble rate $H(z)$. The state finder pairs are defined as 
\begin{eqnarray} 
 j &=&  \frac{\dddot{\mathcal{R}}}{\mathcal{R}H^3}=\frac{\ddot{H}}{H^3}-(2+3q),\\
 s &=& \frac{j-1}{3(q-0.5)}.   
\end{eqnarray}

The $Om(z)$ parameter is defined by
\begin{equation}
Om(z) = \frac{E^2(z)-1}{(1+z)^3-1},
\end{equation}
where $E(z)=\frac{H(z)}{H_0}$ is the dimensionless Hubble parameter. Here $H_0$ is the Hubble rate at the present epoch. If $Om(z)$ becomes a constant quantity, the DE model is considered to be a cosmological constant model with $\omega = -1$. If this parameter increases with $z$ with a positive slope, the model can be a phantom model with $\omega < -1$. For a decreasing $Om(z)$ with negative slope, quintessence model are obtained ($\omega >-1$).

For the hybrid scale factor, the state finder pair become
\begin{eqnarray}
(j,s)&=&\left(\frac{\dddot{\mathcal{R}}}{\mathcal{R}H^3},\frac{j-1}{3(q-0.5)}\right)\nonumber \\
&=&\left(1-\frac{3 \beta}{(\alpha t+\beta)^2}+\frac{2\beta}{(\alpha t+\beta)^3}, \frac{4\beta-6\beta(\alpha t+\beta)}{6\beta (\alpha t+\beta)-9(\alpha t+\beta)^3}\right).
\end{eqnarray} 

The values of the statefinder pair depend on the parameters $\alpha$ and $\beta$. Both $j$ and $s$ evolve with time from large value to small value at late time. At the beginning of cosmic time, the statefinder pair for the HSF are $\{1 + \frac{2-3\beta}{\beta^2}, \frac{2}{3\beta}\}$ whereas at late time of cosmic evolution, the HSF model behaves like $\Lambda$CDM with the statefinder pair having values $\{1, 0\}$. The $Om(z)$ parameter can also be assessed for the HSF. The HSF model behaves like a cosmological constant model for a substantial time zone in the recent past  and before this time zone, the model evolves as a phantom field \cite{Mishra21}.

\section{Some cosmological Models with HSF in GR}
In this section, we review some of the cosmological models with hybrid scale factor in the framework of general relativity. We will set up the field equations with different matter field and discuss the results. 

\subsection{Dark Energy models in anisotropic Bianchi V Space time in GR}
We may consider an anisotropic Bianchi V space time in the form
\begin{equation} \label{ST}
ds^{2}=-dt^{2}+A^{2}dx^{2}+e^{2 a x}(B^{2}dy^{2}+C^{2}dz^{2}),
\end{equation}
where $A=A(t), B=B(t), C=C(t)$ are functions of cosmic time $t$ only and $a$ is a positive constant. The Einstein Field equations can be written as 
\begin{equation} \label{EFE}
G_{ij} \equiv R_{ij}-\frac{1}{2}Rg_{ij}=-\frac{8\pi G}{c^4}T_{ij},
\end{equation}
where $G_{ij}$ Einstein tensor, $R_{ij}$ Ricci tensor, $R$ Ricci scalar, $T_{ij}$ energy momentum tensor, which can also be assumed for multiple fluids. For simplicity, we take $8\pi G=c=1$.

For the anisotropic Universe, we may define the directional Hubble parameters as $H_x=\frac{\dot{A}}{A}$, $H_y=\frac{\dot{B}}{B}$, $H_z=\frac{\dot{C}}{C}$. The mean Hubble parameter can be obtained as $H=\frac{\dot{\mathcal{R}}}{\mathcal{R}}=\frac{1}{3}(H_x+H_y+H_z)$, where $\mathcal{R}$ is the scale factor. The shear scalar $\sigma^2\left[=\sigma_{ij}\sigma^{ij}=\frac{1}{2} \left(\Sigma H^2_i-\frac{\theta^2}{3}\right)\right]$ usually considered to be proportional to the scalar expansion $\theta(=3H)$. This leads to an anisotropy relation between the metric potentials $B$ and $C$ in the form $B=C^k$. Now, the relationship between the scale factor and the metric potentials can be established as $A=\mathcal{R}$, $B=\mathcal{R}^{\frac{2k}{k+1}}$ , $C=\mathcal{R}^{\frac{2}{k+1}}$. Also the Hubble parameter and directional Hubble parameters can be related as $H_x=H$, $H_y=\left(\frac{2k}{k+1}\right)H$ and $H_z=\left(\frac{2}{k+1}\right)H$. It can be noted that the rate of expansion along $x$-axis is same as the mean Hubble parameter. Now, the set of field equations in the form of Hubble parameter become
\begin{eqnarray}
\dot{H}_y+\dot{H}_z+H_y^2 + H_z^2 + H_y H_z-\frac{a^2}{A^2}&=& -\frac{1}{A^2}[T_{11}],\\ \label{EFE6}
\dot{H}_x+\dot{H}_z+H_x^2 + H_z^2 + H_x H_z-\frac{a^2}{A^2}&=& -\frac{1}{B^2e^{2a x}}[T_{22}],\\ \label{EFE7}
\dot{H}_x+\dot{H}_y+H_x^2 + H_y^2 + H_x H_y-\frac{a^2}{A^2}&=& -\frac{1}{C^2e^{2a x}}[T_{33}],\\ \label{EFE8}
H_xH_y+H_yH_z+H_zH_x-3\frac{a^2}{A^2}&=& -[T_{44}]. \label{EFE9}
\end{eqnarray}

The energy conservation for the anisotropic fluid, $T^{ij}_{;j}=0 $, yields
\begin{equation}\label{ECE}
\dot{\rho}+3\rho(\omega+1)H+\rho(\delta H_x+\gamma H_y+\eta H_z)=0,
\end{equation}

In the above equations, an overhead dot on a field variable denotes differentiation with respect to time $t$.

Eqn. \eqref{ECE} can be splitted into two parts: the first one corresponds to the conservation of matter field with equal pressure along all the directions i.e. the deviation free part of \eqref{ECE}  and the second one corresponds to that involving the deviations of equation of state (EoS) parameter:
\begin{equation}\label{ECE1}
\dot{\rho}+3\rho(\omega+1)H=0,
\end{equation}
and
\begin{equation}\label{ECE2}
\rho(\delta H_x+\gamma H_y+\eta H_z)=0.
\end{equation}

It is now certain that, the behaviour of the energy density $\rho$ is controlled by the deviation free part of EoS parameter whereas  the anisotropic pressures along different spatial directions can be obtained from the second part of the conservation equation. From eqn. \eqref{ECE1}, we obtain the energy density as $\rho=\rho_0 \mathcal{R}^{-3(\omega+1)}$, where $\rho_0$ is the value of energy density at the present epoch.

The field equations in an anisotropic space time can be handled by explicitly writing the matter field. We may consider that, the cosmic fluid be consisting of a mixture of Dark energy and usual fluid or it may be embedded in an external magnetic field. Also, we may consider the effect of bulk viscosity on the cosmic dynamics. After choosing the matter field, we may employ the hybrid scale factor to obtain the pressure and energy density of the Universe and study the cosmic dynamics through the dynamical equation of state parameter. In this context, we will consider some of the results of earlier works~\cite{Mishra15,Mishra18,Mishra19}.

\subsubsection{Case:1}
If we assume a cosmic fluid consisting of a mixture of usual matter and some dark energy components, then the right hand side of \eqref{EFE} can be expressed as~\cite{Mishra15}
\begin{eqnarray} 
-T_{ij}^D &=& diag[\rho_D, -p_{Dx}, -p_{Dy},-p_{Dz}]\nonumber \\ \label{EMTD}
          &=& diag[1, -\omega_{Dx}, -\omega_{Dy},-\omega_{Dz}]\rho_D\nonumber\\
          &=&diag[1,-(\omega_D +\delta), -(\omega_D+\gamma), -(\omega_D+\eta)]\rho_D.
\end{eqnarray}

Here $p_D$, $\rho_D$ and $\omega_D$ respectively denote pressure and energy density and DE EoS parameter $(=\frac{p_D}{\rho_D})$. Also, $\delta, \gamma, \eta$ are the skewness parameters that describe the change of pressure in the respective coordinate axis. With an algebraic manipulations in the field equations, the skewness parameters can be obtained alike as
\begin{eqnarray}
\delta &=& -\left(\frac{k-1}{3\rho_D}\right)\left[\frac{k-1}{(k+1)^2}\times F(H)\right],\label{SP1}\\
\gamma &=& \left(\frac{k+5}{6\rho_D}\right)\left[\frac{k-1}{(k+1)^2}\times F(H)\right],\label{SP2}\\
\eta &=& -\left(\frac{5k+1}{6\rho_D}\right)\left[\frac{k-1}{(k+1)^2}\times F(H)\right],\label{SP3}
\end{eqnarray}
where $F(H)=2(\dot{H}+3H^2)$. The functional $F(H)$ for the hybrid scale factor becomes
$$F(t)=\frac{2}{t^2}\left[3\alpha^2t^2+6\alpha \beta t+3\beta^2-\beta\right].$$

The energy conservation equations for DE fluid ($T_{:j}^{ij(D)}$) can be obtained as
\begin{eqnarray}
\dot{\rho}_D+3\rho_D(1+\omega_D)H+\rho_D(\delta H_x+\gamma H_y+\eta H_z)&=&0. \label{ECD}
\end{eqnarray}

The energy density $\rho_D$ and the EoS parameter $\omega_D$ are obtained as
\begin{equation} \label{EDD}
\rho_D = 2\left(\frac{k^2+4k+1}{(k+1)^2}\right)\frac{\dot{\mathcal{R}}^2}{\mathcal{R}^2}-\frac{3a ^2}{\mathcal{R}^2}= 2\left(\frac{k^2+4k+1}{(k+1)^2}\right)\left(\alpha+\frac{\beta}{t}\right)^2-\frac{3a ^2}{e^{2\alpha t}t^{2\beta}},
\end{equation}

\begin{eqnarray}
\omega_D \rho_D&=&-\frac{2}{3}\left(\frac{k^2+4k+1}{(k+1)^2}\right)\left(2\frac{\mathcal{\ddot{R}}}{\mathcal{R}}+\frac{\dot{\mathcal{R}}^2}{\mathcal{R}^2}\right)+\frac{a^2}{\mathcal{R}^2}\nonumber \\
&=&-\frac{2}{3}\left(\frac{k^2+4k+1}{(k+1)^2}\right)\left[-\frac{2\beta}{t^2}+3\left(\alpha+\frac{\beta}{t}\right)^2\right]+\frac{a ^2}{e^{2\alpha t}t^{2\beta}}.\label{EOS}
\end{eqnarray}

The model studied under the above assumptions being more realistic to simulate a cosmic transit that favours a phantom phase at late time. The use of the hybrid scale factor significantly changes the behavior of the cosmic fluid~\cite{Mishra15}.

\subsubsection{Case:2}
The behaviour of the cosmological model with dark energy fluid and viscous matter fluid is another attraction in the study of cosmic expansion. So, the energy momentum tensor and the energy conservation equations for viscous fluid ($T_{:j}^{ij(V)}$) can be obtained as~\cite{Mishra15,Mishra18}
\begin{equation}
-T_{ij}^V = diag[\rho, -\bar{p}, -\bar{p},-\bar{p}],\\ \label{EMTV}
\end{equation}
and 
\begin{equation}
\dot{\rho}+3(\rho+\bar{p})H=0. \\ \label{ECV}
\end{equation}

We get the energy density for the matter field as
\begin{align} \label{EDMV}
\rho=\frac{\rho_{0}}{\left[ e^{\int{H}.dt}\right]^{3(\epsilon +1)}},
\end{align}
where $\rho_0$ is the integration constant or rest energy density of present time.

The energy density $\rho_D$ and the EoS parameter $\omega_D$ are obtained as
\begin{eqnarray} \label{EDDV}
\rho_D &=& 2\left(\frac{k^2+4k+1}{(k+1)^2}\right)\frac{\dot{\mathcal{R}}^2}{\mathcal{R}^2}-\frac{3a ^2}{\mathcal{R}^2}-\rho_{0} \mathcal{R}^{-3(\epsilon + 1)} \\ \nonumber
&=& 2\left(\frac{k^2+4k+1}{(k+1)^2}\right)\left(\alpha+\frac{\beta}{t}\right)^2-\frac{3a ^2}{e^{2\alpha t}t^{2\beta}}-\dfrac{\rho_{0}}{(e^{\alpha t}t^{\beta})^{\frac{3}{2}(k+1)(\epsilon+1)}},
\end{eqnarray}

\begin{eqnarray}
\omega_D \rho_D &=& -\frac{2}{3}\left(\frac{k^2+4k+1}{(k+1)^2}\right)\left(2\frac{\mathcal{\ddot{R}}}{\mathcal{R}}+\frac{\dot{\mathcal{R}}^2}{\mathcal{R}^2}\right)+\frac{a^2}{\mathcal{R}^2}-\epsilon \rho_{0} \mathcal{R}^{-3(\epsilon +1)} \\ \nonumber
&=& -\frac{2}{3}\left(\frac{k^2+4k+1}{(k+1)^2}\right)\left[-\frac{2\beta}{t^2}+3\left(\alpha+\frac{\beta}{t}\right)^2\right]+\frac{a ^2}{e^{2\alpha t}t^{2\beta}}-\dfrac{\epsilon \rho_{0}}{(e^{\alpha t}t^{\beta})^{\frac{3}{2}(k+1)(\epsilon+1)}}.\label{EOSV}
\end{eqnarray}

The skewness parameters can be obtained as in eqns. \eqref{SP1},\eqref{SP2},\eqref{SP3}. Both the fluids the viscous and and dark energy fluid have shown their dominance respectively in early time and late time of evolution in the presumed hybrid scale factor~\cite{Mishra19}.  

\subsubsection{Case:3}
When the cosmic fluid contains some one dimensional cosmic string aligned along the $x-$ axis along with DE, we have~\cite{Mishra18}
\begin{equation}
-T_{ij}^S = diag[0,-\lambda,0,0]. \label{EMTS}
\end{equation}

For this choice of the matter field, the skewness parameters can be obtained as 
\begin{eqnarray}
\delta &=& -\left(\frac{k-1}{3\rho_D}\right)\left[\frac{k-1}{(k+1)^2}\times F(H)+\lambda]\right],\label{SP4}\\
\gamma &=& \left(\frac{k+5}{6\rho_D}\right)\left[\frac{k-1}{(k+1)^2}\times F(H) +\lambda]\right] ,\label{SP5}\\
\eta &=& -\left(\frac{5k+1}{6\rho_D}\right)\left[\frac{k-1}{(k+1)^2}\times F(H)-\lambda\right]. \label{SP6}
\end{eqnarray}

The energy conservation equation yields
\begin{equation} \label{ECES}
\dot{\rho}+3\left(\rho+p+\frac{\lambda}{3}\right)H=0,
\end{equation}
which can be integrated to obtain
\begin{equation}
\rho=\rho_0 \mathcal{R}^{-3(1+\omega+\xi)},
\end{equation}
where $\rho_0$ is rest energy density due to the matter field at the present epoch. $\xi$ and $\omega$ are assumed to be non evolving state parameters. The equations of state for strings and isotropic fluid can be considered respectively as
\begin{equation}
\lambda=3\xi\rho,~~~~~~~~~~~~~~~~~~~~~~~~~p=\omega\rho.
\end{equation}

Consequently, the pressure and string tension density are obtained as
\begin{equation} 
p=\omega\rho_o \mathcal{R}^{-3(1+\omega+\xi)}~~~~~~~~~~  \lambda=3\xi\rho_0 \mathcal{R}^{-3(1+\omega+\xi)}.
\end{equation}

From Eqs. \eqref{EFE9}, \eqref{EMTD} and \eqref{EMTS}, we can obtain the DE density as
\begin{equation}
\rho_D=3(\Omega_{\sigma}-\Omega_k)\left(\frac{\dot{\mathcal{R}}}{R}\right)^2-\rho,
\end{equation}
where $\Omega_{\sigma}=1-\frac{\mathcal{A}}{2}$, $\Omega_k=\frac{a^2}{\mathcal{R}^2}$. 

The density parameters can be expressed as
\begin{equation}
\Omega_D=\Omega_{\sigma}-\Omega_k-\Omega_m,
\end{equation}
where $\Omega_m=\frac{\rho_m}{3H^2}$.

The total density parameter becomes
\begin{equation} \label{TDP}
\Omega=\Omega_m+\Omega_D=\Omega_{\sigma}-\Omega_k.
\end{equation}

The DE EoS parameter $\omega_D$ is now obtained as
\begin{equation}
\omega_D \rho_D=- \left[2\frac{\ddot{\mathcal{R}}}{\mathcal{R}}+\left(\frac{\dot{\mathcal{R}}}{\mathcal{R}}\right)^2\right]\Omega_{\sigma}+\frac{a^2}{\mathcal{R}^2}-\rho(\omega+\xi).
\end{equation}

The model constructed with the hybrid scale factor in the work~\cite{Mishra18}, yields anisotropy in the dark energy pressure that evolves with the cosmic expansion at least at late times. However, at an early time, the dynamics of the accelerating universe substantially affected by the presence of cosmic string.

\subsubsection{Case:4}
We may extend our investigation further by incorporating an external electromagnetic field in both $x$~\cite{Mishra19a} and $z$~\cite{Ray19} direction with the dark energy fluid. Here, we will incorporate the electromagnetic field in the form  
\begin{equation}\label{EMTE1} 
E_{ij} = \frac{1}{4 \pi} \left[ g^{sp}f_{\mu s}f_{\nu p}-\frac{1}{4} g_{\mu \nu}f_{sp}f^{sp} \right],
\end{equation}
where $g_{\mu \nu}$ is the gravitational metric potential and $f_{\mu \nu}$ is the electromagnetic field tensor. In order to avoid the interference of electric field, we have considered an infinite electrical conductivity to construct the cosmological model. This results in the expression $f_{14}=f_{24}=f_{34}=0$. Again, quantizing the axis of the magnetic field along $x$-direction as the axis of symmetry, we obtain the expression $f_{12}=f_{13}=0,$ $f_{23}\neq 0$. Thus, the only non-vanishing component of electromagnetic field tensor is $f_{23}$. With the help of Maxwell's equation, the non-vanishing component can be represented as, $f_{23}=-f_{32}= k_1 $, where  $k_1$ is assumed to be a constant and it comes from the axial magnetic field distribution. For the anisotropic BV model, the components of EMT for electromagnetic field can be obtained as
\begin{equation}\label{EMTE2}
E_{ij}= diag[-1,A^2, - B^2 e^{2 \alpha x}, - C^2 e^{2 \alpha x}]\mathcal{M},
\end{equation}
where $\mathcal{M}=\frac{k_1^{2}}{B^{2}C^{2}e^{4 \alpha x}}$. When $k_1$ is non-zero, the presence of magnetic field has been established along the $x$-direction. If $k_1$ vanishes, the model will reduce to the one with DE components only. Hence
\begin{eqnarray}
\delta &=& -\left(\frac{k-1}{3\rho_D}\right)\left[\frac{k-1}{(k+1)^2}\times F(H)+2\mathcal{M}\right],\label{eqn:2.1.7}\\
\gamma &=& \left(\frac{k+5}{6\rho_D}\right)\left[\frac{k-1}{(k+1)^2}\times F(H) +2\mathcal{M}\right] ,\label{eqn:2.1.8}\\
\eta &=& -\left(\frac{5k+1}{6\rho_D}\right)\left[\frac{k-1}{(k+1)^2}\times F(H)-2\mathcal{M} \right].\label{eq:2.1.9}
\end{eqnarray}

The energy conservation equations for magnetized field can be be obtained as
\begin{equation} \label{ECEE}
\dot{\mathcal{M}}+2 \mathcal{M} (H_{y}+H_{z})=0.
\end{equation}

The DE energy density $\rho_D$ and the EoS parameter $\omega_D$ are obtained as
\begin{eqnarray} 
\rho_D &=& 2\left(\frac{k^2+4k+1}{(k+1)^2}\right) \dfrac{\dot{\mathcal{R}}^2}{\mathcal{R}^2}- 3 \dfrac{\alpha ^2}{\mathcal{R}^2}+ 2\mathcal{M},\\
\omega _D\rho_D &=& -\frac{2}{3}\left(\frac{k^2+4k+1}{(k+1)^2}\right)\left[\dfrac{2\ddot{\mathcal{R}}}{\mathcal{R}}+ \dfrac{\dot{\mathcal{R}}^2}{\mathcal{R}^2} \right] + \dfrac{\alpha ^{2}}{\mathcal{R}^2}-\frac{2}{3}\mathcal{M}.
\end{eqnarray}

The findings of the model indicate that at late phase of cosmic evolution, the universe on its expansion is in agreement with the supernovae observation and in the deceleration phase, it allows the formation of large scale structure of the Universe. The presence of magnetic field changes the dynamics substantially at least at the initial phase of the evolution. The skewness parameters are more sensitive to the magnetic field strength and the choice of the model parameters.

\section{Conclusion}
In the present review, we have discussed the role of a hybrid scale factor in obtaining viable cosmic dynamics without assuming any specific relationship between the pressure and energy density of the Universe. The hybrid scale factor is an intermediate between the power law and exponential law expansion behaviour. The HSF has two factors: one behaving like the power law expansion and dominates at an early phase of cosmic evolution. The other factor behaves like an exponential expansion and dominates at late phase to provide a model closer to the concordance $\Lambda$CDM model at late times. Also, the observations supporting the late time cosmic speed up phenomena suggest that, the Universe has undergone a transition from an early deceleration to late time acceleration phase. Such a cosmic transit behaviour envisages a signature flipping behaviour of the deceleration parameter. The usual power law or exponential law scale factors generate constant deceleration parameter and are therefore not suitable for cosmological studies relevant to cosmic speed up issue. In this context, a hybrid scale factor can simulate a deceleration parameter with early positive values and late time negative values.  However, the HSF has two adjustable parameters which need to be tuned so as to obtain viable models with cosmic transit behaviour. Out of the two parameters, one can be readily constrained to some acceptable range, but the other parameter may be constrained through a detailed analysis of the $H(z)$ data. In this context, an accuracy in the $H(z)$ data is essential to get models closer to the present Universe. We have considered some cosmological examples, where, hybrid scale factor have been used to obtain satisfactorily the cosmic dynamics. It is shown that, the HSF can be good alternative in providing interesting and viable models.

\section*{Acknowledgement} SKT, BM and SR thank IUCAA, Pune, India for providing support through the visiting associateship program. The work by MK has been performed with a support of the Ministry of Science and Higher Education of the Russian Federation, Project "Fundamental problems of cosmic rays and dark matter", No 0723-2020-0040.

\end{document}